\documentclass[a4paper,10pt]{article}

\usepackage{url}

\usepackage{graphicx}  %
\usepackage{amsmath}
\usepackage{bm}  %
\usepackage{tikz}
\usepackage{epsfig}
\usepackage{times}
\usepackage{color}
\usepackage{hyperref}
\usepackage[section]{placeins}
\usepackage{todonotes}
\usepackage{caption}
\newcommand{\etal}{et al.}
 
\renewcommand{\Re}{\operatorname{Re}}

\newcommand{\doi}[1]{\href{http://doi.org/#1}{doi:#1}}

\newcommand*{\prv}{Phys. Rev. } 
\newcommand*{\pra}{Phys. Rev. A } 
\newcommand*{\prc}{Phys. Rev. C } 
\newcommand*{\anphy}{Ann. Phys. }
\newcommand*{\ppnp}{Prog. Part. Nucl. Phys. }
\newcommand*{\inpreparation}{\textit{in preparation}}

\begin{document}
%
\title{Universality and the Coulomb interaction}
%
%
\author{Christiane~H.~Schmickler$^{1,2}$}
%
%
%

\maketitle              

\noindent$^1$Institut f\"ur Kernphysik, Technische Universit\"at Darmstadt,
64289\ Darmstadt, Germany\\
schmickler@theorie.ikp.physik.tu-darmstadt.de,\\
$^2$RIKEN Nishina Center, RIKEN, Saitama 351-0198, Japan

\begin{abstract}
We study the relationship between universal effects and 
the Coulomb interaction. Here, we present our approach and 
a first illustrative example.

Understanding the relationship of universality and the Coulomb 
interaction is important for weakly bound nuclear few-body systems. 
The example nucleus we study is the excited state of $^{17}$F which 
we model as a proton and an $^{16}$O core. We use a Gaussian potential 
to represent short-range forces together with a Coulomb potential. 
Our calculation uses the Gaussian Expansion Method (GEM), which is 
a variational method well-suited to our problem. 
We find that we need to choose the range of the Gaussian potential 
in addition to the potential strength that reproduces the right 
scattering length. We propose to fit the range of the Gaussian potential 
such that the effective range is reproduced. We show that this 
approach leads to consistent results for $^{17}$F. 

\end{abstract}
\section{Introduction}

The Efimov effect has been in the focus of few-body research ever since 
its observation in cold atom experiments \cite{kraemermark2006}. Even before 
that, there was discussion in the literature about whether and where 
the Efimov effect plays a role in nuclear physics \cite{efimov1991}. 

However, the long-range Coulomb interaction complicates this picture in the 
nuclear sector and many studies to date focussed on the question of whether 
there would be an Efimov state without the Coulomb force. 

In our studies, of which this paper is a first excerpt, we try to answer 
the question of which effect the Coulomb force has on would-be Efimov states 
and under which circumstances the Efimov picture is relevant in spite of the 
Coulomb interaction. 

To this end, we use well-tested methods of calculating universal properties 
of few-body systems, namely Gaussian potentials and variational methods, 
and add a Coulomb interaction to the system.

Here, we investigate a specific example system, $^{17}$F$(\frac{1}{2}^+)$, 
which we describe as an $^{16}$O nucleus plus a single proton. This state 
is a proton halo and a shallow bound state. As a dimer it is well-suited 
as a first test of our methods. The trimer and tetramer calculations will 
be presented in an upcoming more in-depth article.

\section{Interaction and Methods}
As we are interested in universal behaviour, we use a simple Gaussian potential plus
the Coulomb potential. 

\begin{equation}
 H_2 = -\frac{\hbar^2}{2\mu}\frac{\partial^2}{\partial r^2} +  V_0 e^{-\frac{r^2}{2r_0^2}} + \hbar cZ_1Z_2\frac{\alpha}{r}
\end{equation}

We solved this system with the Gaussian Expansion Method (GEM) that was 
developed by Kamimura \etal \cite{hiyamakino2003}. It is a variational 
method with geometrically spaced Gaussian basis functions and it is well-suited to this 
problem because the matrix elements can be calculated analytically and it 
gives accurate results without much fine tuning.

\section{Coulomb Modified Effective Range Expansion}
The usual effective range expansion has to be modified to 
account for the fact that the Coulomb interaction is long-range. 
This is achieved by simply replacing the free outside wave functions 
used to calculate the scattering length and effective range 
by Coulomb functions. 

This leads to the Coulomb-modified effective range expansion:
\begin{equation}
 C^2_{\eta,0}p\cot\tilde\delta_0(p) + \gamma h(\eta) = -\frac{1}{a_C}+\frac{1}{2}r^{C}_\text{eff}p^2 + \dots,
 \label{eq:coulombmodifiedeffrangeexp}
\end{equation}
where $p$ is the wave number, $\eta = \mu c^2\alpha Z_1Z_2/(\hbar cp)$ and  $\gamma = 2\frac{\mu c^2}{\hbar c}\alpha Z_1Z_2$ 
with the reduced mass $\mu$ and charge numbers $Z_1$ and $Z_2$ , $C^2_{\eta,p} = \frac{2\pi\eta}{e^{2\pi\eta}-1}$ , $h(\eta) = \Re \frac{\Gamma'(i\eta)}{\Gamma(i\eta)}-\log(\eta)$ and 
$\tilde\delta_0(p)$ is the phase shift between the ingoing and outgoing Coulomb wave functions.
This determines the Coulomb-modified scattering length $a_C$ and the Coulomb-modified effective 
range $r^C_\text{eff}$. For more detail, the reader is referred to \cite{koenigphd}\cite{bethe1949}%
\cite{vanhaeringenkok1982}.

\section{Application to \texorpdfstring{$^{17}$F}{17F}}

We will concentrate on showing one application, namely $^{17}$F as a dimer of 
$^{16}$O and a proton,
in this paper and refer to another publication for more general results 
and more applications \cite{inpreparation}.

The first excited state of $^{17}$F is a well-known proton halo. 
This means that in first approximation it can be described as a 
dimer of the core ($^{16}$O) and a proton \cite{rybergforssen2016}. From the values for the 
binding momentum $\gamma = 13.6\,\text{MeV}$ and the ANC $A \approx 80\,\text{fm}^{-\frac{1}{2}}$ presented in 
\cite{rybergforssen2016}, we calculated the $^{16}$O-p Coulomb modified scattering length
and the Coulomb modified effective range to be $a_C = 4475.96\,\text{fm}$ and 
$r_\text{eff}^C = 1.18737\,\text{fm}$ according to the formulae derived in \cite{rybergforssen2016}. 

\begin{figure}[tb]
 \bigskip
\begin{tikzpicture}[scale=0.5]
 \node at (-0.3,0) {\includegraphics[height=3.75cm]{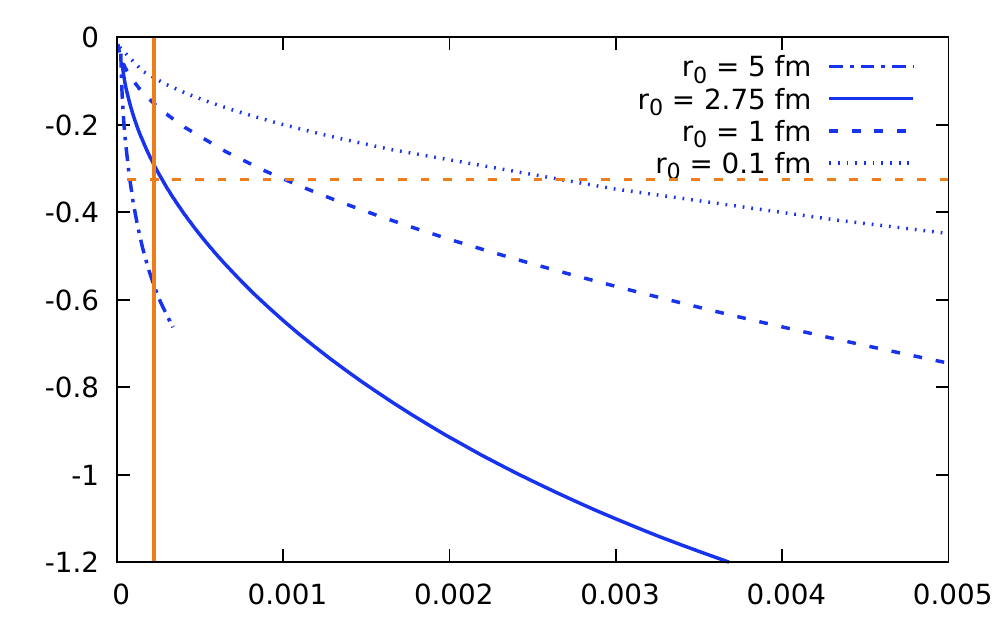}};
 \node[rotate=90, anchor=north] at (-6.9,0.3) {$\sqrt{E}$ [MeV$^{\frac{1}{2}}$]};
 \node at (-0.3,-4.2){$\frac{1}{a_C}$ [fm$^{-1}$]};

 \node at (12,3.6) {\includegraphics[width=6cm,trim=0 0 0 165, clip]{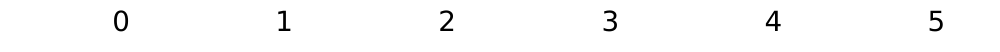}};
  \node at (12,0) {\includegraphics[height=3.75cm]{reffvsr0beiasapprox4500fv2.pdf}};
  \node at (14.3,-1.4) {\colorbox{white}{\includegraphics[width=3.6cm]{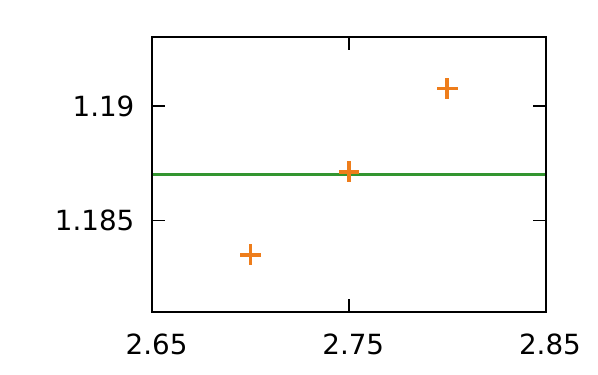}}};
  \node[rotate=90, anchor=north] at (5.5,0) { $r_{\text{eff}}^C \quad [\text{fm}]$};
  \node at (10.6, -4.2)  { $r_0 \quad [\text{fm}]$};
\end{tikzpicture}
\caption{\emph{Left panel:} Different dimer predictions for different values of the range $r_0$ of the Gaussian short-range potential. 
 The vertical orange line is the value of the scattering length in the $^{16}$O+p system, the horizontal orange line 
 is the binding energy of the $^{17}$F($\frac{1}{2}^+$) state. \emph{Right panel:} Determination of the effective range at the point of the $^{16}\text{O-p}$ scattering length for different $r_0$. 
 The horizontal line shows the value of the Coulomb modified effective range in the $^{16}$O+p system.}
 \label{fig}
\end{figure}

However, plugging in the parameters for this system, i.e. $a_C$, $Z_1 = 1$, $Z_2=8$ and the 
reduced mass $\mu \approx 873\,\text{MeV}$,
will lead to an ambiguity because  we 
obtain different curves for different $r_0$ as shown in 
the left panel of Fig.~\ref{fig}.
This ambiguity has to be resolved in order to make any kind of prediction. From the point of view 
of the effective range expansion, Eq.~(\ref{eq:coulombmodifiedeffrangeexp}), the most obvious parameter to fit after the scattering length is the 
effective range.

To this end, we calculated the effective range for different ranges $r_0$ of the Gaussian short-range potential.
We selected results reproducing the scattering length of $a_C = 4475.96\,\text{fm} \pm 80\,\text{fm}$. These are 
shown in the right panel of Fig.~\ref{fig}. The value of $r^C_\text{eff}$ rises monotonically with $r_0$ in the range we explored. 
Thus, we can uniquely identify a value for $r_0$ that reproduces the correct Coulomb modified effective range for this 
system. For $^{16}$O+p, $r_0 = 2.75\,\text{fm}$.

Going back to the left panel of Fig.~\ref{fig}, we can see that the determined value for $r_0$ reproduces the binding energy of the 
$^{17}$F proton halo state well. The physical value is at the point where the horizontal and vertical lines cross, 
and the curve for $r_0 = 2.75\,\text{fm}$ crosses the vertical line very close to that point.  

Note that this is only a consistency check, because in the formula 
for the scattering length from \cite{rybergforssen2016} the effective range and the binding energy is an input. 
The effective range in turn contains as additional input the asymptotic normalization coefficient, which can be measured. 
Therefore we use the binding energy as an input to determine the scattering length, which we in turn use as 
an input for obtaining the binding energy. Since the methods that were used to connect scattering length and binding energy 
are different, however, this is a good consistency check between the methods.

\section{Summary and Conclusion}

We showed in this paper a first demonstration of our method, 
i.e. using a Gaussian potential together with the Coulomb interaction
to test universal properties of weakly bound states. We 
applied our method to a proton halo state, $^{17}$F$(\frac{1}{2}^+)$, which 
we regarded as a dimer of a proton and a $^{16}$O core. 

To make the result unique we have to fix $r_0$, which is the 
range of our Gaussian potential. This is unnecessary in systems
without the Coulomb interaction, because the result would be the 
same for all $r_0$ after converting it into the right units. 
Having to fix $r_0$ is due to the Coulomb interaction 
introducing an additional scale to the system. 

We chose to fix $r_0$ such that the Coulomb-modified effective 
range would be reproduced, which gave us the correct binding energy 
for $^{17}$F. This is a promising first test of our method 
which we can build upon in future work. 

We will expand this approach to trimers and tetramers 
of identical charged bosons in an upcoming publication. 

\section{Acknowledgments}
The author thanks H.-W. Hammer for a careful reading of the 
manuscript and 
helpful comments and E. Hiyama for teaching her the 
Gaussian Expansion Method. 

Funded by the Deutsche Forschungsgemeinschaft (DFG, German Research Foundation) - Projektnummer 279384907 - SFB 1245.


\end{document}